\documentclass[runningheads]{llncs}
\usepackage{graphicx}
\usepackage{soul}
\usepackage{booktabs}
\usepackage{xcolor}
\usepackage[linesnumbered,algoruled,lined, boxed]{algorithm2e}

\begin{document}
\title{Designing Limitless Path in Virtual Reality Environment \thanks{Accepted in International HCI 2021 (24-29 July 2021) - http://2021.hci.international}}
%
%
\author{Raghav Mittal\and
Sai Anirudh Karre \and
Y. Raghu Reddy}
\authorrunning{M. Raghav et al.}

\institute{
Software Engineering Research Center\\ IIIT Hyderabad, India\\
\email{\{raghav.mittal,saianirudh.karri\}@research.iiit.ac.in}
\email{raghu.reddy@iiit.ac.in}
}

\maketitle              
\begin{abstract}
Walking in a Virtual Environment is a bounded task. It is challenging for a subject to navigate a large virtual environment designed in a limited physical space.  External hardware support may be required to achieve such an act in a concise physical area without compromising navigation and virtual scene rendering quality. This paper proposes an algorithmic approach to let a subject navigate a limitless virtual environment within a limited physical space with no additional external hardware support apart from the the regular Head-Mounted-Device (HMD) itself. As part of our work, we developed a Virtual Art Gallery as a use-case to validate our algorithm. We conducted a simple user-study to gather feedback from the participants to evaluate the ease of locomotion of the application. The results showed that our algorithm could generate limitless paths of our use-case under predefined conditions and can be extended to other use-cases. 
\keywords{Limitless Paths; Bounded Virtual Environment; Virtual Reality Products}
\end{abstract}

\section{Motivation}
Virtual Reality (VR) environments are designed from a participant's perspective. In HMDs, participant is considered to be the center of the VR scene and the environment is oriented around the participant. This idea of centrality helps define the boundary of the VR environment for a given scene.  VR practitioners normally design a full-scale scene to orient the participant within a fixed bounded environment rather than building a dynamic bounded environment due to various reasons. Some of the reasons like HMD limitations, physical space limits, dependency on additional hardware support, poor scene baking, poor frame rate, etc. influence the VR practitioners to build navigation controls through hand-held devices by letting the participant stay stationary. This contradicts the idea of creating realness in VR scene as the participant is forced to navigate the scene through hand-held controllers but in reality s(he) is stationary. To help address this issue, we developed an approach to let the participant navigate beyond the control of a hand-held device by taking into account the following two factors:

\begin{itemize}
    \item \textbf let the participant physically navigate in the VR environment without the influence of external haptic hardware.
    \item \textbf let the participant navigate seamlessly with a limitless path in a limited virtual play area.
\end{itemize}

The above two factors can be addressed by considering the underlying technical aspects
\begin{itemize}
    \item As part of the predefined constraints, physical environment and virtual environment ratio must be maintained
    \item An algorithm needs to be built to generate a limitless path with no obstruction for the participant to navigate in a virtual environment setup in a limited room space.
    \item VR environment assets need to be generated based on the path and orientation of the participant in the virtual environment
    \item VR environment should appear infinite to the participant. However, physically the participant will still be navigating in a limited predefined space.
\end{itemize}

\begin{figure}
    \centering
    \includegraphics[scale=0.5]{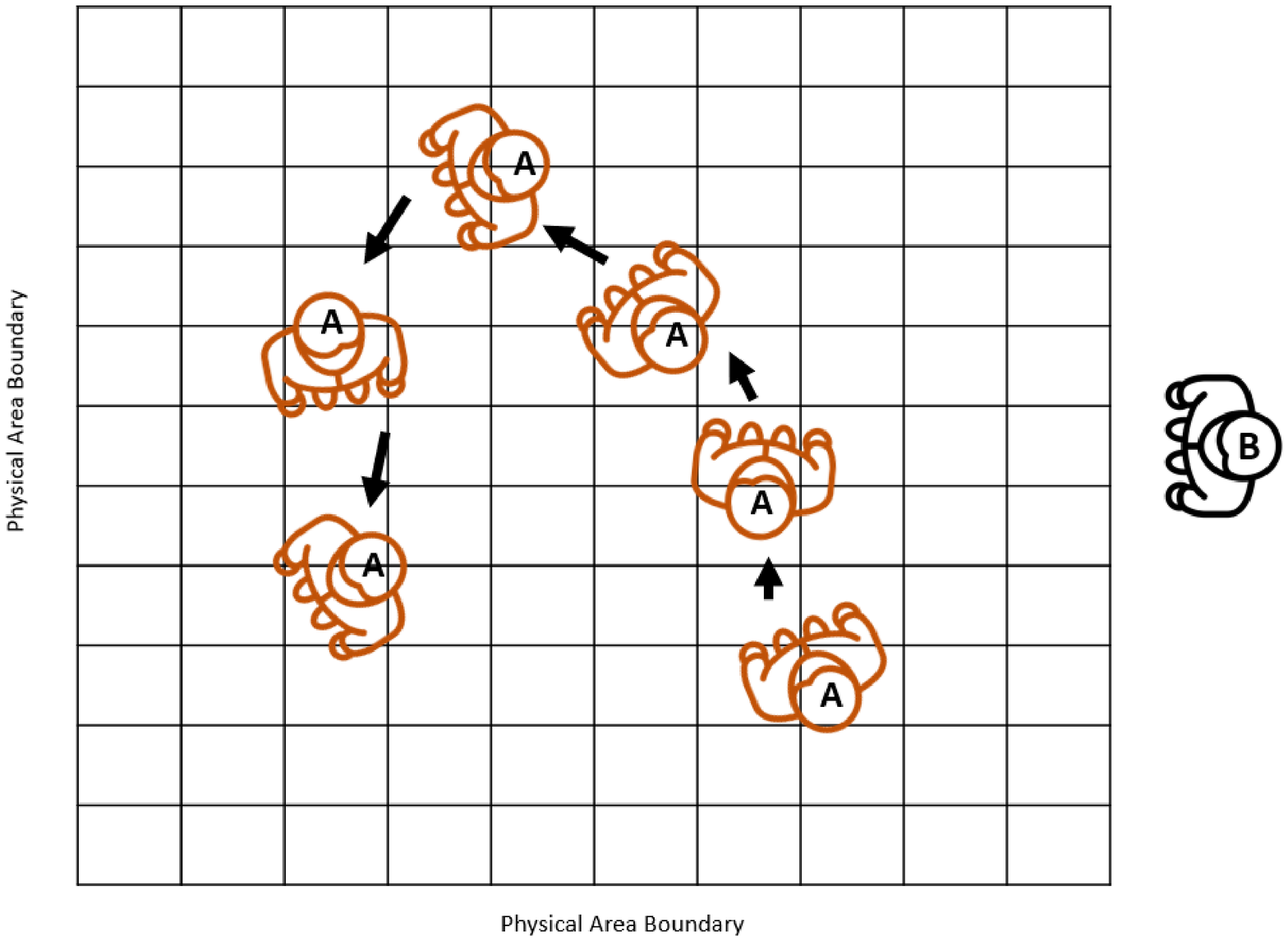}
    \caption{Person A experiencing limitless path in a finite physical environment}
    \label{fig:personexperience}
\end{figure}

In this work, we use the term \textbf{\textit{`Limitless Path'}} in the context of obstruction-free navigation in a VR environment within a limited physical environment. We define limitless path as a programmatically generated never ending path in a VR scene. This path is limitless and unbounded in terms of length. The progress of the path is automatic and the scene assets are generated inline with the generated path. For illustration consider Fig \ref{fig:personexperience},  person A - standing in a 10ft x 10ft physical grid wearing a HMD and another person B outside this physical grid. Person A loads a VR scene with the support of limitless path implementation and infinitely walks in this limited space. Person A experiences an endless walk in the VR scene. For person B,  person A appears to be walking within a 10ft x 10ft physical grid area continuously in a random path within the limited physical area.

The rest of the paper is structured as follows: In section 2, we detail  our approach towards designing and implementing path generation and boundary detection in a VR environment. In Section 3, we present the user study on a small set of participants. In section 4, we discuss the threats to validity. In section 5, we present some related work in this area and finally in section 6 we present some conclusions.

\section{Our Approach}
Continuous walking within the specified area (with in the scope of a VR scene or a physical area) is a critical aspect of path generation, as any generated path must be within the specified boundaries of the VR scene. Thus, path generation and boundary detection plays a key role in the  progress of the participant along the limitless path. As part of our work, we designed and implemented both the path generation and boundary detection algorithms in a virtual environment. 

\begin{figure}[h]
    \centering
    \includegraphics[scale=0.8]{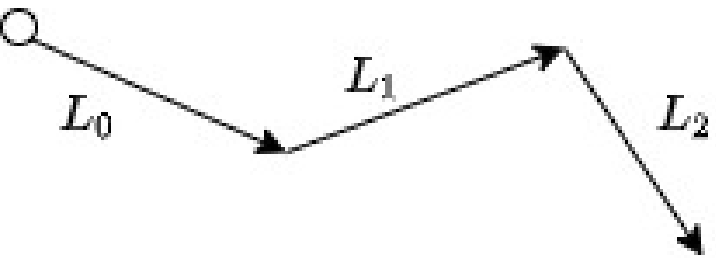}
    \caption{Path generated on start of system. \(L_0\) starts from player's position. }
    \label{fig:examplelinesegment}
\end{figure}

\begin{figure}[h]
    \centering
    \includegraphics[scale=0.8]{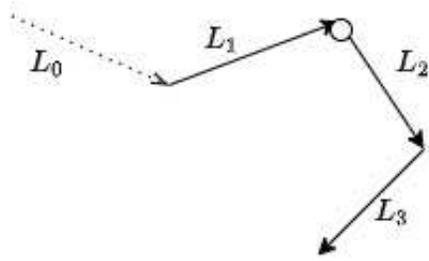}
    \caption{\(L_0\) is removed and \(L_3\) is added when the player reached to the end of \(L_1\).}
    \label{fig:linesegmentcontinue}
\end{figure}

Consider a use-case where a path has to be presented to the participant in a particular VR context. At the time of the VR scene generation,  we are aware of the start position of the participant with in a virtual environment. As the participant walks forward, the path is generated by  addition and removal of line segments in the path to create a perception of continuity. This path generation system outputs the path as a line made of multiple connected line segments, as shown in Fig \ref{fig:examplelinesegment}. The designed system will generate 3 successive line segments (shown in Fig. \ref{fig:examplelinesegment} as \(L_0\),\(L_1\), and \(L_2\)) from the starting point. When the player reaches the end of \(L_1\), the first segment \(L_0\) is removed, and a new segment \(L_3\) is added to the end of \(L_2\) as shown in Fig \ref{fig:linesegmentcontinue}. This process keeps repeating until the participant terminates the program. 

Each generated path line segment forms an obtuse angle with the previous line segment in most cases. This is done to ensure comfortable locomotion to the participant by avoiding sharp turns. It also reduces the possibility of the path intersection or path override when a new line segment is added to the path.

\subsection{Palling - Path Generation}
In this section, we provide step-by-step details of our path generation algorithm. We term \textit{`Palling'} as a user action to move forward in the virtual environment for ease of terms. \textit{1 Pal} unit is equal to 1 unit distance taken by the user in the virtual environment. For purposes of simplicity, we assume the dimension of the rectangular bounded area from $(0, 0)$ to $(x, z)$. If the user starts at $0, 0)$, the user can take \textit{z Pal} units to reach $0, z)$ and \textit{x Pal} units to reach $x, 0)$. In order to generate a path, below are the necessary inputs:

\begin{itemize}
    \item Player's starting position $_0$and head-yaw $\beta_0$at that position.
    \item Dimensions of the boundary $(x, z)$, here \textit{x} and \textit{z} are the limits of the boundary in a plane along \textit{x-axis} and \textit{z-axis}.
    \item Path properties include segment length and path width i.e the perpendicular distance between parallel boundaries representing the width of the path in virtual environment.
\end{itemize}

\begin{figure}[h]
    \centering
    \includegraphics[scale=0.8]{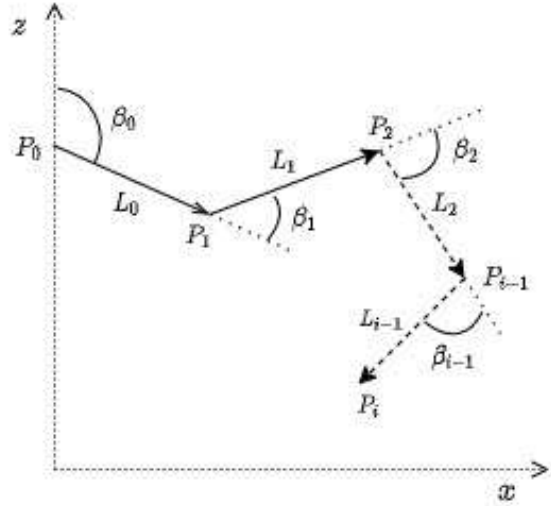}
    \caption{Locomotion of a participant in the play area}
    \label{fig:locomotion_playarea}
\end{figure}

As shown in Fig \ref{fig:locomotion_playarea}, upon start of the application, the first line segment $L_0$ has a starting point $P_0$ and terminal point $P_1$ i.e. it is represented as $L_0$ = $\overline{P_0P_1}$. Similarly, for $L_1$, $P_1$ and $P_2$ are starting and terminal points respectively. Eq \ref{one} generalizes this for $L_i$. Currently the length is considered to be a constant value and provided as a necessary input. We plan on computing this automatically depending on the boundary area in future.

\begin{equation} \label{one}
L_i = \overline{P_i P_{i+1}}
\end{equation}

In the given play area, the coordinates of position \(P_{i+1}\) are calculated as shown in following equations \ref{two} and \ref{three}, where \(0<=i<\) total number of segments generated since the start of the application. Here \(\beta_0\) is head-yaw (HMD's orientation along y-axis in virtual environment) of the user.
\begin{equation} \label{two}
P_{i+1}(x) = l * \sin{\beta_{i}} + P_{i}(x) 
\end{equation}
\begin{equation} \label{three}
P_{i+1}(z) = l * \cos{\beta_{i}} + P_{i}(z)
\end{equation}

When we recursively run equations \ref{two} and \ref{three}, we generate a limitless path in a defined boundary of \(D(x, z)\). The generated path is instantaneous, nonlinear, and limited by certain Pal units due to the bounded area. The upcoming \(P_i\) needs to be generated by detecting the proximity of $P_{i-1}$ to the boundary. When $P_{i-1}$ is not close to the boundary, \(\beta_i\) is set to a random value in range \(\{\beta_{i-1}-\pi/2, \beta_{i-1}+\pi/2\}\). The procedure of generating the next line segment \(L_i\) for a given position \(P_i\) at a boundary is defined on \(\beta_i\) value.

\subsection{Pragging - Boundary Detection}
In this section, we detail our boundary detection algorithm. For ease of terms, we refer to a user instance on detecting a boundary or hitting a boundary as \textit{`Pragging'}. \textit{1 Prag} unit is equal to one hit at a boundary. 

\begin{figure}
    \centering
    \includegraphics[scale=0.7]{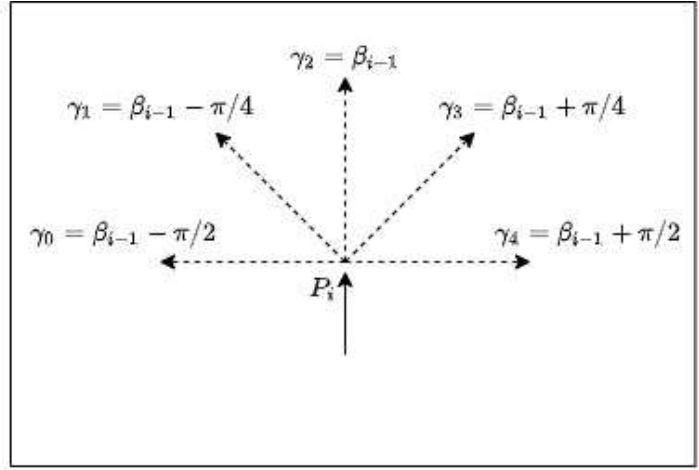}
    \caption{Dashed lines represent rays, solid line represents the direction of the line segment \(L_{i-1}\) of the existing path. Here \(j=4\).}
    \label{fig:rayprojections}
\end{figure}

To ensure that the player doesn’t cross the boundary of the given application, the generated path must not span beyond the bounded-area. To achieve this, the bounded-area is enclosed into wall-like colliders. We define \textit{`j'} as a value that equally divides a 180$^{\circ}$ range into multiple possible rays as shown in Fig \ref{fig:rayprojections}. Whenever a new point $P_{i+1}$ has to be generated, \textit{j+1} number of rays are projected in multiple directions with certain angles called as $\gamma$ where \(k\) is in range \(\{0, j\}\) and \(j>0\). The range of angle here is \(\beta_{i-1} - \pi/2, \beta_{i-1} + \pi/2\)
\begin{equation}
    \gamma_j = \beta_{i-1} - \pi/2 + ((\pi/j) * k)
\end{equation}

If the value of j=4, we have j+1 rays i.e. 5 rays at equal angles between \(\beta_{i-1} - \pi/2, \beta_{i-1} + \pi/2\) are called $\gamma_0,  \gamma_1, \gamma_2, \gamma_3, \gamma_4$ as shown in Fig \ref{fig:rayprojections}. Here the source of the rays is \(P_i\) and length is equal to \(path\_length + path\_width/2\) 

\begin{figure}
    \centering
    \includegraphics[scale=0.8]{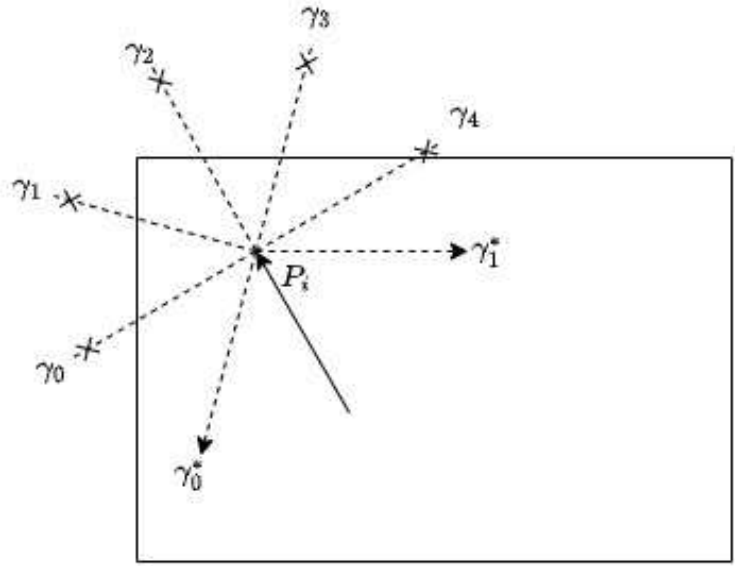}
    \caption{When all the rays $\gamma_0$....$\gamma_{j+1}$ hit the boundaries then two more rays are generated to decide the direction of $L_i$.}
    \label{fig:boundary-case-1}
\end{figure}

\begin{figure}
    \centering
    \includegraphics[scale=0.8]{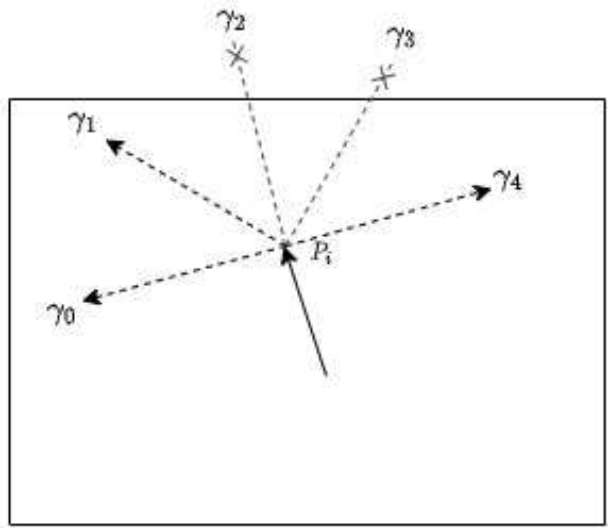}
    \caption{The \(\gamma\) for next \(L_i\) will be chosen from $\gamma_0$, $\gamma_1$, $\gamma_4$.}
    \label{fig:boundary-case-2}
\end{figure}

If a ray collides with the play area boundary,  we term it as 1 Prag unit. If \textit{j=a}, we have \textit{a+1} rays generated. Out of these \textit{a+1} rays, one of the $\gamma_i$ direction is chosen to generate a path $L_i$. Below are the possible cases in which the generated rays can collide with a boundary:

\begin{itemize}
    \item If $\gamma$ = \textit{a+1 Prag }units, i.e. all generated rays hit the boundary as shown in Fig \ref{fig:boundary-case-1}. This means that the $P_i$ is at the corner of the bounded area. In this case, $\pm135^{\circ}$ is the way to go away from the boundary. Two more rays are shot in directions \(\gamma^{*}_0= \beta_{i-1} - 3\pi/4\) and \(\gamma^{*}_1=\beta_{i-1} + 3\pi/4\) to avoid corner as an escape strategy. The ray which don't hit the boundary is chosen as the direction for $L_i$. If $P_i$ is equidistant from boundaries then one of the ray from $\gamma^{*}_0$ and $\gamma^{*}_1$ is chosen randomly.
    
    \item If $1\leq\gamma\leq a\ Prag\ units$ i.e. not all but atleast one ray hits the boundary as shown in Fig \ref{fig:boundary-case-2} and anyone of the non-hitting rays can be used to generate $L_i$.
    \item If $\gamma$ = \textit{0 Prag} units, i.e. none of the rays hit any of the boundaries. Here the path is free from boundaries as shown in Fig \ref{fig:rayprojections}. Then the \(\beta_i\) is randomly chosen from range \({\beta_{i-1}-\pi/2, \beta_{i-1}+\pi/2}\) for generating the upcoming $L_i$.
\end{itemize}


\subsection{PragPal Algorithm}
As part of our work, we conduct pragging and palling simultaneously to generate a limitless path for a user in a virtual environment. Algorithm \ref{algo} provides the pseudo-code of our concept, which is used in the simulation \footnote{https://github.com/raghavmittal101/path\_gen\_sys/}. 

\begin{algorithm}[H] \label{algo}
\caption{PragPal Algorithm Overview}
\SetKwInput{KwInput}{Input}                
\SetKwInput{KwOutput}{Output}              
\DontPrintSemicolon
  
  \KwInput{user position, head yaw, line segment length, path width}
  \KwOutput{List of 2D points x,z which represent a path}

  \SetKwFunction{FMain}{Main}
  \SetKwFunction{FPathGen}{GeneratePoint\_Pal}
  \SetKwFunction{FBetaRange}{GenerateBeta\_Prag}
 
  \SetKwProg{Fn}{Function}{:}{}
  \Fn{\FPathGen{$beta$, $point$}}{
        \textit{point.x} = \textit{segment\_length}*sin(\textit{beta}) + \textit{point.x}\;
        \textit{point.z} = \textit{segment\_length}*cos(\textit{beta}) + \textit{point.z}\;
        \KwRet point
  }

    \SetKwProg{Fn}{Function}{:}{}
    \Fn{\FBetaRange{$beta$, $point$}}{
        j = 4 // can be any value greater than 1\;
        valid\_directions = []\;
        for k in range (0, j+1): // k = 0, 1, 2, 3, 4\;
        \hskip1.5em ray\_direction = \textit{beta} - $\pi/2 + (\pi/j) * k)$\;
        
        \hskip1.5em ray = Generate ray in direction ray\_direction\ from \textit{point}\;
        \hskip1.5em if ray do not hit boundary:\;
        \hskip1.5em\hskip1.5em valid\_directions.push(ray\_direction)\;
        if valid\_directions.size\textless 0:\;
        \hskip1.5em beta = randomly pick element from valid\_directions\;
        else:\;
        \hskip1.5em \textit{beta} = randomly choose from range \{$beta-\pi/2, beta+\pi/2$\}\;
        
        \KwRet beta \;
    }
    
  \SetKwProg{Fn}{Function}{:}{\KwRet}
  \Fn{\FMain}{
        point = current position of player: [x, z]\;
        beta = current head-yaw of player in radians\;
        points\_list = [] // points represent the shape of current path \;
        points\_list.Append(point)\;
        \textbf{OnSceneStart:} // called only on initialization of the scene\;
        \hskip1.5em for i in range(4):\;
        \hskip1.5em\hskip1.5em point = GeneratePoint\_Pal(beta, point) \;
        \hskip1.5em\hskip1.5em points\_list.Append(point)\;    
        \hskip1.5em\hskip1.5em beta = GenerateBeta\_Prag(beta, point)\;
        
        \textbf{OnSceneUpdate:} // called whenever player reaches end of segment\;
        \hskip1.5em points\_list.pop()  // removes first element\;
        \hskip1.5em point = GeneratePoint\_Pal(beta, point)\;
        \hskip1.5em beta = GenerateBeta\_Prag(beta, point)\;
        \hskip1.5em points\_list.Append(point)\;
        
  }
\end{algorithm}

\subsection{Prototype Implementation of Virtual art gallery}
To understand our work's effectiveness, we implemented the algorithm to build a VR based 'Virtual Art Gallery'. It is an endless corridor composed of two walls running parallel to each other. The user can walk through the gallery to explore the art items displayed on these walls. When the user progresses in the forward direction, i.e., palls forward, using the tracked Pal units, the path generation logic updates the upcoming path and auto-generates the corridor with relevant assets in the VR environment.  

The corridor consists of two wall-like 2D mesh structures positioned parallel to the path. Whenever the new path is updated, the assets related to the corridor are also updated. In our example application, a set of royalty-free floral drawing images are displayed in the gallery. As the user navigates in the scene, Prag units are simultaneously tracked to detect boundaries to generate the possible paths. A collider trigger placed at the end of each line segment in the current path captures the Prag units.  Whenever the user reaches a turn, this method is triggered. Thus generating an infinite limitless path for the participant in VR.

\begin{figure}[h]
    \centering
    \includegraphics[scale=0.7]{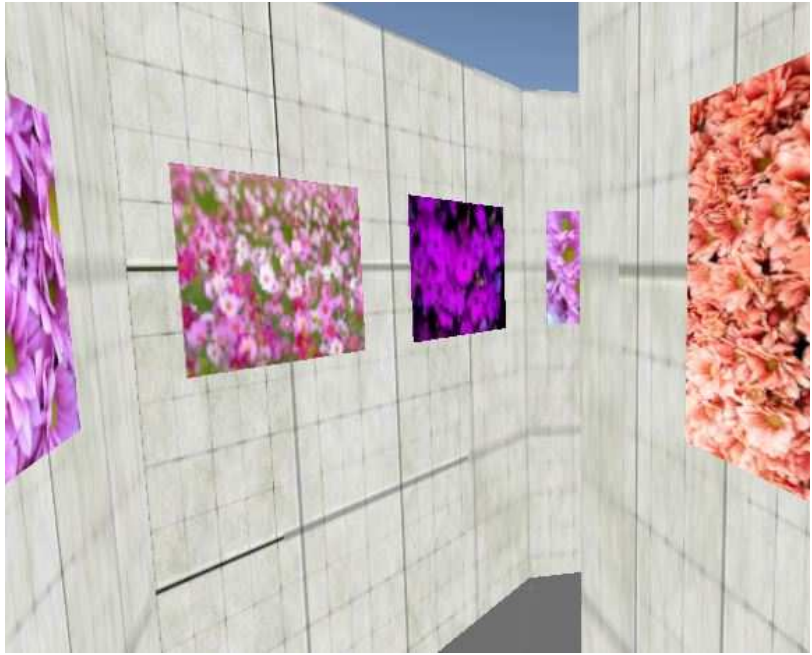}
    \caption{Inside view the virtual art gallery.}
    \label{fig:sceneView1}
\end{figure}

\begin{figure}[h]
    \centering
    \includegraphics[scale=0.8]{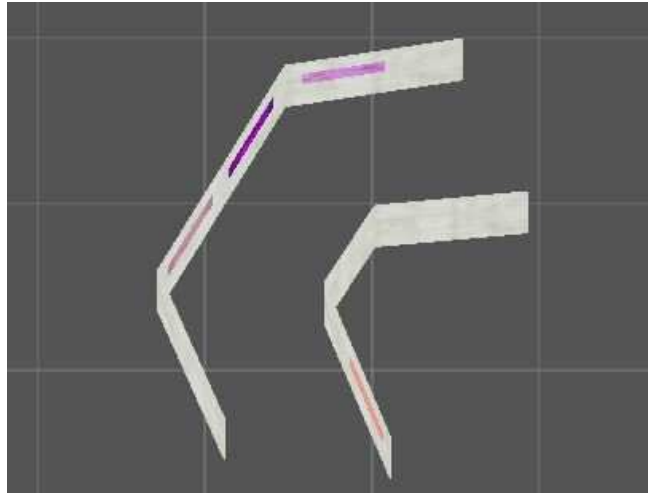}
    \caption{Top view of the virtual art gallery.}
    \label{fig:sceneView2}
\end{figure}

We developed this use-case in Unity3D 2019 LTS, and tested it using Oculus Quest 2019 HMD. The images in the scene are placed at the player's standing height. This is to provide a comfortable view of the player during locomotion. Dimensions of each image are set to 0.5 x 0.5 units in Unity3D (~50 cm x 50 cm). A margin of 0.1 units was given on both sides of each image to keep the walls less cluttered as shown in \ref{fig:sceneView1}. The number of images placed on each wall varied because all the walls cannot necessarily be of the same length, as shown in   figure \ref{fig:sceneView2}. Here, the maximum perpendicular distance between the walls called \textit{path\_width} is always maintained. This was set to 1.2 meter to have enough space for comfortable locomotion. The maximum length of a segment \textit{segment\_length} of path was set to 1.3 meter. Physical space availability is an important factor to consider while deciding the values of path\_width and segment\_length. A larger space can accommodate a broader and longer corridor. 

\subsection{Corridor Generation}
\begin{figure}[h]
    \centering
    \includegraphics[scale=0.6]{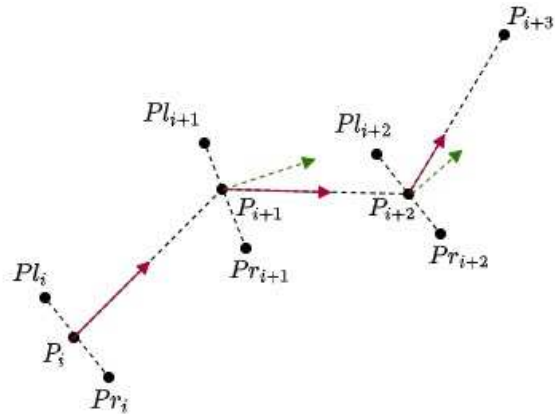}
    \caption{Placement of left and right points for corridor generation.}
    \label{fig:corridorPoints}
\end{figure}

The Pragpal algorithm generates $P_i$, $P_{i+1}$, $P_{i+2}$ ..... and so on. It outputs points that can form a line to be used by a participant to traverse the limitless path in a virtual environment. In order to generate a corridor for walking, as shown in Fig. \ref{fig:sceneView1} and Fig. \ref{fig:sceneView2}, we generate points $Pr_i$ and $Pl_i$ on right and left side of each point $P_i$ respectively. These two new points $Pr_i$ and $Pl_i$ are equidistant from $P_i$ and the distance between $P_i$  and $Pr_i$ or $Pl_i$ is half the path width ($path\_width$). In the current version of our approach, the $path\_width$ is a static value defined by the programmer and is set as a default width of the visible or imaginary corridor. However, the path width may be changed depending the size of the boundary and future iterations of our work shall accomodate dynamic generation of path-width based on the VR scene and bounded area size. As $P_i$, $P_{i+1}$ are generated by PragPal algorithm, the red arrow in Fig \ref{fig:corridorPoints} from $P_i$ is normal vector to $Pr_i$ and $Pl_i$ towards $P_{i+1}$. Once we reach $P_{i+1}$, we move towards $P_{i+2}$ along the red arrow. However, to position $Pr_{i+1}$ and $Pl_{i+1}$, we generate a average vector (shown as green dotted arrow) of $\overrightarrow{P_{i+1} - P_i}$ vector and $\overrightarrow{P_{i+2} - P_{i+1}}$. The resultant average vector is dotted green arrow. Using this average vector, we generate $Pr_{i+1}$ and $Pl_{i+1}$. Similarly, $Pr_{i+2}$ and $Pl_{i+2}$ and $Pr_{i+3}$ and $Pl_{i+3}$ so on will be created along with PragPal points. Joining all $Pr_i$'s and $Pl_i$'s points will provide us the corridor path required for our virtual art gallery. The code-implementation of this corridor generation\footnote{https://github.com/SebLague/Curve-Editor/tree/master/Episode\%2006} is included as part of our virtual art gallery scene. Using this corridor generation method, we generate boundary to the path width for better path visualization. 

This use-case is a simple instance of validating the limitless path approach. Our approach can be applied to various other use-cases or domains like entertainment, scientific research, studying spatial cognition, education, etc. to validate limitless paths.

\section{User Experience Study}
This section contains details about the user experience study conducted to understand our implementation's caveats, immersion, and ease-of-locomotion. Our use-case simulation is available on GitHub for practitioners for further experimentation \footnote{https://github.com/raghavmittal101/path\_gen\_sys/} along with its visualization for practitioners' analysis \cite{resources}.
\\
\newline
\textbf{Experiment Setup}- We have set up a physical room with a scale of 24 feet x 17 feet at our university campus to run the virtual art gallery use-case. We used Oculus Quest HMD, a standalone virtual reality headset with inside-out tracking capability. It helps determine the subject's position and orientation with no additional hardware apart from the HMD itself. Its technical specifications include a display resolution of 1440 x 1600 pixels per eye, 72Hz refresh rate, Qualcomm Snapdragon 835 processor, and 4GB RAM. As part of our VR scene, a set of 20 images are presented in the virtual art gallery. For test purposes, these images are set to be repeated till the termination of the environment. Due to the Coronavirus pandemic, we conducted our tests under COVID-19 protocol and  limited the number of users who participated in the study. Special care was taken to sanitize the apparatus after each trial and follow. VR face-masks were provided to the participants to avoid direct face contact with the HMD headset.
\\
\newline
\textbf{Participants} - All the subjects who participated in this study were university students/staff recruited randomly. Among the participants, there were six males and four females. The mean age was 24.5 years. Before undergoing the study, we asked them a few questions to understand their prior exposure to VR. We observed that 35\% of the participants had used  video/computer games regularly, and 29\% had prior VR experience.
\\
\newline
\textbf{Tasks} - After obtaining participant consent and educating them about VR induced sickness, we asked them to complete the following tasks in the given order:
\begin{itemize}
    \item \textbf{Demographic Survey}\footnote{https://forms.gle/CW5kAWjAz7oTsyb26}: Participants were asked fill out data regarding Age, Sex, Gaming experience, and VR experience.
    \item \textbf{Exploration Task}: Participants had to locomote in the virtual art gallery VR Scene for about 5 minutes at their will.  Post 5 minutes, the scene was terminated externally by the experimenter by alerting the participant verbally.
   \item \textbf{Questionnaires}: Participants were requested to fill Simulator Sickness Questionnaire(SSQ) \cite{10.1207/s15327108ijap0303} to understand simulation sickness while navigating the algorithm generated path, Igroup Presence Questionnaire(IPQ) \cite{igroup} to understand the participant presence experience in the scene generated by the algorithm.
\end{itemize}

\begin{table}[h]
\centering
\caption{Mean and Weighted \% of User Experience of Virtual Art Gallery}
\label{res}
\begin{tabular}{@{}lll@{}}
\toprule
Factors & Mean & \%Weighted Response \\ \midrule
Nausea & 1.3 & 6.19\% \\
Oculomotor & 3.2 & 15.2\% \\
Disorientation & 2.8 & 13.3\% \\
Sense of being & 5.2 & 74.2\% \\
Spatial Presence & 25.9 & 74\% \\
Involvement & 18.7 & 66\% \\
Experienced Realism & 14.8 & 52\% \\
Presence & 64.6 & 65\% \\ \bottomrule
\end{tabular}
\end{table}

\noindent \textbf{Results and Observations} - Table \ref{res} provides the results of our survey study. We observe that the subject has a good understanding of spatial presence and sense of being in the virtual reality scene despite automatically generating the subject's path in the virtual environment. The subjects experienced some disorientation and minor Nausea levels while being part of our VR scene automated path generation. There is a reasonable amount of involvement of participant while navigating in the VR scene. 

\section{Threats to Validity}
\textbf{Internal Validity} - We conducted a controlled experiment of our approach using a limited setup. We incrementally recorded the experiment's updates and conducted a minimal study in an optimal setup consisting of a rectangular physcial room. We reviewed the study design with fellow researchers and gathered feedback.  Our physical experiment results match the results of simulation studies conducted through automated means. However, given the covid restrictions, the interactions were fairly limited and virtual in nature. A more realistic in-person review of the study design could have yielded better insights.
\newline
\\
\textbf{External Validity} - We made every attempt to conduct a simplified study of a use-case among the few participants. Our results show that our approach can be easily extended to a large population. However, a serious user-study with a large sample size is required to understand the underlying challenges of our approach. Additionally, creating different types of rooms in a iterative manner could have provided a different set of results.   
\newline
\\
\textbf{Construct Validity} - We coined Palling and Pragging as units to determine the motion and halt of a user in the virtual environment for ease of terms. It helped us establish the distance traveled versus the number of times the user hit the boundary. Using these two new units, we could ascertain the user's course and progress in a virtual environment.
\\
\newline
\textbf{Conclusion Validity} - In our study, we don't claim our approach  as optimal compared to existing methods. There is a possibility of more effective ways of limitless path generation. A comparative study among all available methods is essential to determine the statistical significance of our approach over others.

\section{Related Work}
Williams et al. conducted experiments on understanding the differences between locomotion in virtual environments using Joystick and locomotion through physical turning \cite{10.1145/1140491.1140495}. Their work is oriented towards understanding the spatial limitations through the different means of locomotion only. They found that large physical spaces are required for comfortable locomotion. A small space limits the length of the generated path. Such a problem can be solved by scaling up the translational gain or developing alternate methods of locomotion. Darken et al. \cite{darken1997omni} are the first to develop an omnidirectional treadmill, and Souman et al. have extended it to a Cyberwalk Omnidirectional treadmill to enable infinite locomotion in a finite space \cite{10.1145/2043603.2043607}. This setup requires external haptic hardware support and may not provide the appropriate end user experience in the virtual environment for locomotion. Juyoung et al. have discussed the differences in treadmill based Walk-in-place methods, and Non-treadmill based Walk-in-place methods \cite{10.1145/3343055.3361926}. Griffin et al. compare the locomotion techniques, which require a hand-held device for locomotion. For instance, teleportation versus the hands-free techniques like walk-in-place \cite{10.1145/3242671.3242707}. They observed that hands-free techniques offer a higher presence than hand-busy techniques.

Sun et al. proposed a saccadic redirected walking technique, which worked based on eye-tracking. It allows infinite walking in a room-scale environment without the need for hardware like ODTs. However, one of the major downsides of it was that it required HMD with eye-tracker and high processing resources. \cite{10.1145/3197517.3201294}. Matsumoto et al. employed a visual-haptic approach to let the user walk on an unlimited virtual straight line. Their method required installing a circular ring-like structure around which the user was required to walk while feeling the boundary with one hand \cite{10.1145/3359997.3365705}. Such setup makes the overall HMD setup immobile, expensive and less adaptable to the masses. Vasylevska et al. proposed flexible spaces technique in which the environment consisted of 4 rooms with partially overlapping areas, but the overlap was not visible to the user. Such an environment creates a perception of limitlessness because the user cannot see beyond the wall of their current room \cite{1db93dc337774f4ba35adc67e58049fd}. Such an approach can be tweaked to provide multiple path options to the participants.  Suma et al. extended this work with an argument of maximizing walking in a limited space. Their algorithm dynamically modifies the layout of the scene, which consists of two overlapping rooms. Their overlapping is dynamically modified throughout the game-play to create an illusion of infiniteness, but not a true space \cite{6165136}.

Rietzler et al. proposed a redirection approach in which the user walks in a radius of 1.5 meters. They tried to address the motion sickness caused due to the mismatch between vestibular and visual inputs. According to their work, when walking in a circular path of radius more than or equal to 1.5 meters, the users cannot differentiate between walking in a circular path, or a straight line \cite{10.1145/3313831.3376821}. This was a unique observation and can be used for future advancements on non-haptic-based walking in the VR environment. Conventionally, teleportation methods do not require the user to walk, and hence it is not good for presence. However, Liu et al. presented a teleportation method in which a portal to a location is generated when the user points towards a location to teleport. The user has to walk and step into the portal to teleport. Here the portals are defined in a single room-scale boundary \cite{10.1145/3242587.3242601}. Thus providing limitless locomotion through teleportation.

Ruddle et al. compared the performance of participants in an object searching task done in 4 different conditions. They include search in the real-world, search in the virtual environment through a screen display, search in VE through an HMD with controllers and search in VE with real locomotion. They found the performance to be the most accurate in VE with real locomotion when compared with real-world performance \cite{10.1145/1502800.1502805}. They observed that the natural locomotion interface is the most suitable interface for locomotion in virtual environments because it provides translational and rotational body-based information. Most of these studies are either haptic based or illusion based to achieve a limitless path.  

\section{Conclusion}
In this paper, we presented an approach to implement limitless path generation in Virtual environments using our PragPal Algorithm. We discussed path generation and boundary detection methods used as part of the algorithm. We implemented the algorithm on a Virtual Art Gallery use-case to understand the robustness through simulation. We also conducted a small user study to capture feedback from real-world participants. Our results and observations are promising and provide reasonable motivation to proceed with our planned work. We are currently working on a comparative study through a systematic literature review to understand the merits and demerits of all available locomotion methods in regards to limitlessness. 

As part of future work, we plan to conduct a large scale use-study to understand presence, accommodation, and ease of use experiences from real-world participants to make our algorithm flexible for other implementations. We plan to tweak our algorithm to enable flexibility with the input parameters, for example, length of the line segment, boundary size, width of the path, etc.  

In the current approach, only a single path is generated. The participant has to take that path and can only move in the forward direction. Given that the prior line segments are removed in the process of forward movement, retracing the steps is not possible. We plan on extending the algorithm to facilitate create of a multi-path environment. This will have its own set of challenges, especially from a rendering and overlaying perspective when the participant can move either forward in different directions or retrace their steps to a certain point. We also plan to introduce assets in the use cases that involve interaction between the participant and the asset. This will facilitate development of more use-cases towards realizing limitless path navigation for recreation, phobia studies, banking and other education related applications that involve interaction for task completion.

\bibliographystyle{splncs04}
\bibliography{my-bib}
%





\end{document}